\documentclass[12]{article}

\usepackage{amssymb,latexsym,amsmath}     
\usepackage{dsfont}
\usepackage{upgreek}
\usepackage{graphics}
\usepackage{graphicx}
\usepackage{lscape}
\usepackage{multirow}
\usepackage{amssymb}
\usepackage{dsfont}
\usepackage{authblk}
\usepackage[switch,columnwise]{lineno}
\begin{document}

\title{On Lindley-Exponential Distribution: Properties and Application}
\author[1]{Deepesh Bhati}
\author[2]{Mohd. Aamir Malik}
\affil[1] {Assistant Professor, Department of Statistics, Central University of Rajasthan, deepesh.bhati@curaj.ac.in}
\affil[2] {Department of Statistics, Central University of Rajasthan
aamirmalik.stats@gmail.com}
\maketitle
\begin{abstract}
In this paper, we introduce a new distribution generated  by Lindley random variable which offers a more flexible model for modelling lifetime data. Various statistical properties like distribution function, survival function, moments, entropy, and limiting distribution of extreme order statistics are established . Inference for a random sample from the proposed distribution is investigated and maximum likelihood estimation method is used for estimating parameters of this distribution. The applicability of the proposed distribution is shown through real data sets. 
\end{abstract}



\section{Introduction}
\indent Lifetime distribution represents an attempt to describe, mathematically, the length of the life of a system or a device. Lifetime distributions are most frequently used in the fields like medicine, engineering etc. Many parametric models such as exponential, gamma, Weibull have been frequently used in statistical literature to analyze lifetime data. But there is no clear motivation for the gamma and Weibull distributions. They only have  more general mathematical closed form than the exponential distribution with one additional parameter.\\
\indent Recently, one parameter Lindley distribution has attracted the researchers for its use in modelling lifetime data, and it has been observed in several papers that this distribution has performed excellently. The Lindley distribution was originally proposed by Lindley \cite{11} in the context of Bayesian statistics, as a counter example of fudicial statistics which can be seen that as a mixture of exp($\theta$) and {gamma(2, $\theta$)}. More details on the Lindley distribution can be found in Ghitany et al. \cite{7}. \\
\indent A random variable X is said to have Lindley distribution with parameter $\theta$ if its probability density function is defined as:\\
\begin{equation} \label{1}
f_X(x;\theta)=\frac{\theta^2}{(\theta+1)}(1+x) e^{-\theta x} ;x > 0,\theta>0
\end{equation}
with cumulative distribution function
\begin{equation*}
F(x)=1-\frac{e^{-\theta x}(1+\theta+\theta x)}{1+\theta}
\end{equation*} 
\indent Some of the advances in the literature of Lindley distribution are given by Ghitany et al. \cite{5} who has introduced a two-parameter weighted Lindley distribution and has pointed that Lindley distribution is particularly useful in modelling biological data from mortality studies. Mahmoudi et. al. \cite{12} have proposed generalized Poisson Lindley distribution. Bakouch et al. \cite{3} have come up with extended Lindley (EL) distribution, Adamidis and Loukas \cite{1} have introduced exponential geometric (EG) distribution. Shanker et. al. \cite{14} have introduced a two-parameter Lindley distribution. Zakerzadeh et al.\cite{15} have proposed a new two parameter lifetime distribution: model and properties. M.K. Hassan \cite{7} has introduced convolution of Lindley distribution. Ghitany et al.\cite{6} worked on the estimation of the reliability of a stress-strength system from power Lindley distribution.  Elbatal et al.\cite{4} has proposed a new generalized Lindley distribution.\\
\noindent Risti\'c \cite{13} has introduced a new family of distributions with survival function given by 
\begin{equation*}
\bar F_X(x)= \frac{1}{\Gamma(\alpha)}\int\limits_{0}^{-\log\left(G(x) \right)} t^{\alpha-1}e^-t dt, \quad \quad x\in \mathds{R}, \theta > 0
\end{equation*}

\indent In this paper we introduce a new family of distribution generated by a random variable $ T $ which follows one parameter Lindley distribution. The survival function of this new family is given as:
\begin{equation} \label{2}
\bar F_X(x)= \frac{\theta^2}{1+\theta}\int\limits_{0}^{-\log(G(x))} (1+t)e^{-\theta t} dt, \quad \quad x\in \mathds{R}, \theta > 0
\end{equation}
where $ \theta > 0 $ and ${G}(x)$ is a cumulative distribution function(cdf) which we use to generate a new distribution. The cdf $ G(x) $ is referred to as a transformer and the corresponding probability density function (pdf) is given by
\begin{equation} \label{3}
f(x)=\frac{\theta^2}{1+\theta}\left(1-\log G(x)\right) G(x)^{\theta-1} g(x),
\end{equation}
We consider the transformer to follow exponential distribution with cdf  $G(x)=(1- e^{-\lambda x})$. Hence the survival function of the new distribution is given by
\begin{equation} \label{4}
\bar{F}(x)=1-\frac{ \left(1-e^{-\lambda x}\right)^{\theta}\left( 1+\theta- \theta \log\left(1-e^{-\lambda x} \right)\right)}{1+\theta}
\end{equation}  
with corresponding density given by
\begin{equation}   \label{5}
f(x)=\frac{ \theta ^2 \lambda e^{-\lambda  x} \left(1-e^{-\lambda x}\right)^{ \theta-1} \left(1-\log \left(1-e^{-\lambda x}\right)\right)}{(1+\theta)}
\end{equation}
We refer the random variable with survival function (4) as Lindley-Exponential(L-E) distribution with parameters $\theta$ and $\lambda$  which we denote by L-E($\theta,\lambda$).

The aim of this paper is to study the mathematical properties of the L-E distribution and to illustrate its applicability. The contents are organized as follows. The analytical shapes of the pdfin equations (5) are established in section 2. The quantile function presented in section 3. The expressions for the moment generating function and moments corresponding to Equation (5) are given in Section 4. Limiting distribution of sample statistics like maximum and minimum has been shown in section 5. In section 6, entropy of L-E distribution is presented. The maximum likelihood estimation procedure is  considered in Section 7. The performance of the maximum likelihood estimators for small samples is assessed by simulation in Section 8.  Section 9 gives estimation of stress-strength parameter R by using maximum likelihood estimation method. Finally we conclude the paper by showing applicability of the model to the real data sets.

\section{Shape of the density}
Here, the shape of pdf (5) follows from theorem 1. \\

\noindent \textbf{Theorem 1:} The probability density function of the L-E distribution is decreasing for $ 0<\theta<1 $ and unimodel for $\theta>1$. In the latter case, mode is a root of the following equation:
\begin{equation*}
(1-\theta e^{-\lambda x})\left(1-\log(1-e^{-\lambda x})\right)+e^{-\lambda x}=0
\end{equation*}
\noindent \textit{Proof:} The first order derivative of $ \log(f(x)) $ is 
\begin{equation}  \label{6}
\frac{d \log(f(x))}{d x}=\frac{\lambda \, r(x)}{\left(1-e^{-\lambda x}\right) \left(1-\log \left(1-e^{-\lambda x}\right)\right)}
\end{equation}
where, $r(x)=\left(\theta  e^{-\lambda x}+\left(1-\theta  e^{-\lambda x}\right) \log \left(1-e^{-\lambda x}\right)-e^{-\lambda x}-1\right)$. For $0<\theta<1$, the function $r(x)$ is negative. So $f'(x)<0$  for all $x>0$. This implies that $f$ is decreasing for $0< \theta <1$. Also note that, $(\log f)'(0)=\infty$ and $(\log f)'(\infty)<0$. This implies that for $\theta>1$, $ g(x) $ has a unique mode at $x_0$ such that $r(x) >0$ for $x<x_0$ and $r(x)<0$ for $x>x_0$. So, $g$ is unimodal function with mode at $x=x_0$. The pdf for various values of $\lambda$ and $\theta$ are shown in Figure 1.
\begin{flushright}
$ \Box $
\end{flushright}
\begin{figure}
\centering
\includegraphics[width=0.9\textwidth]{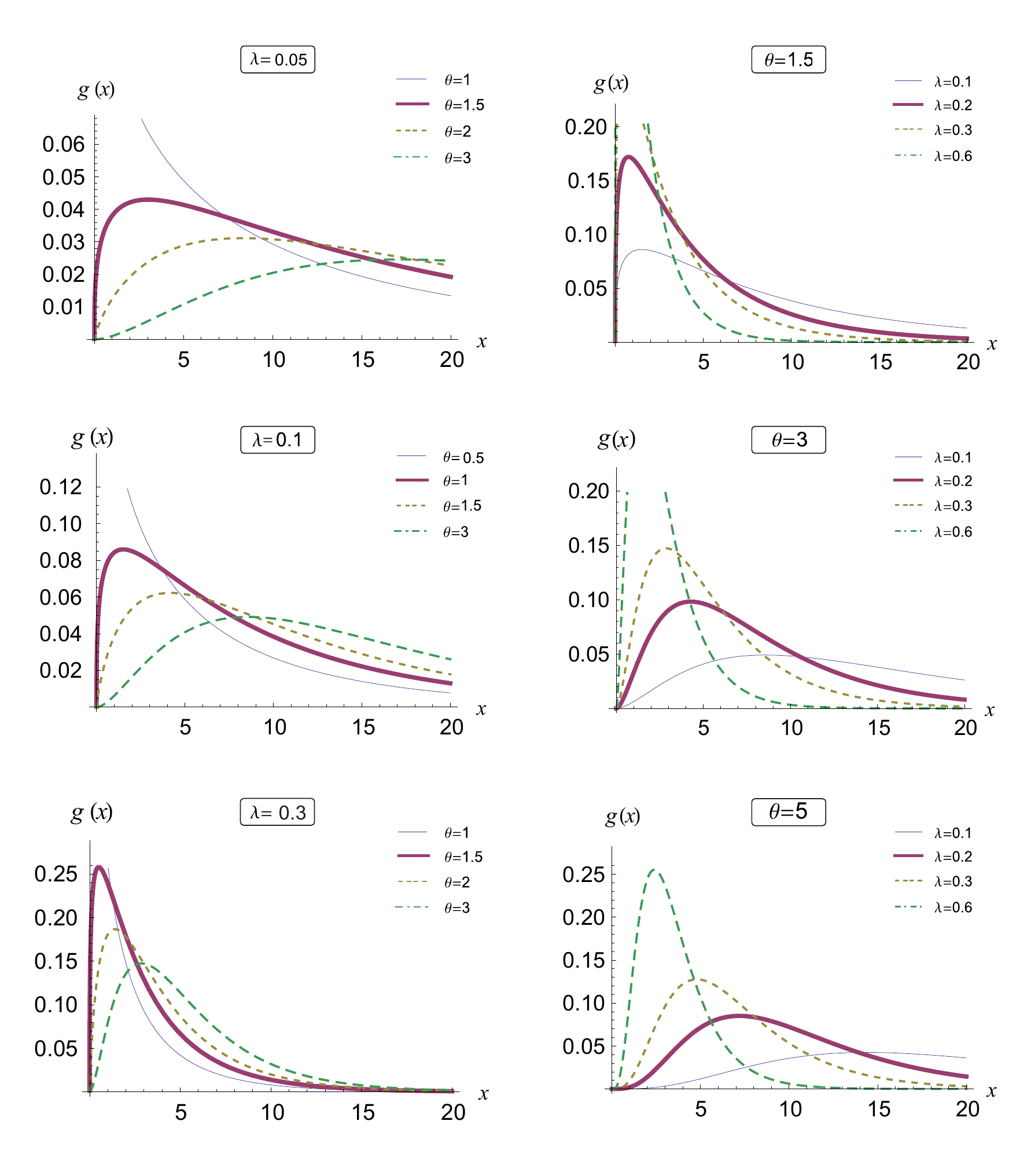}
\caption{PDF plot for various values of $\lambda$ and $\theta$.}
\end{figure}
\noindent We, now,  consider the hazard rate function (hrf) of the L-E distribution, which is given by
\begin{equation}  \label{7}
h(x)=\frac{\theta ^2 \lambda  e^{-\lambda  x} \left(1-e^{-\lambda x}\right)^{\theta-1} \left(1-\log \left(1-e^{-\lambda x}\right)\right)}{1+\theta -\left(1-e^{-\lambda x}\right)^{\theta } \left(1+\theta -\theta \log\left(1-e^{-\lambda x}\right)\right)}
\end{equation}
\\
\noindent \textbf{Proposition 1}: For $\theta >0$ the hazard rate function follows relation $\lim\limits_{x \rightarrow \infty} h(x) = \lambda$. \\

\noindent \textbf{{Proof:}} The proof is straight forward and is omitted. \\

\noindent In Figure 2, hazard function for different values of parameters $ \theta $ and $ \lambda $.
\begin{flushright}
$ \Box $
\end{flushright}
\begin{figure}
\centering
\includegraphics[width=0.9\textwidth]{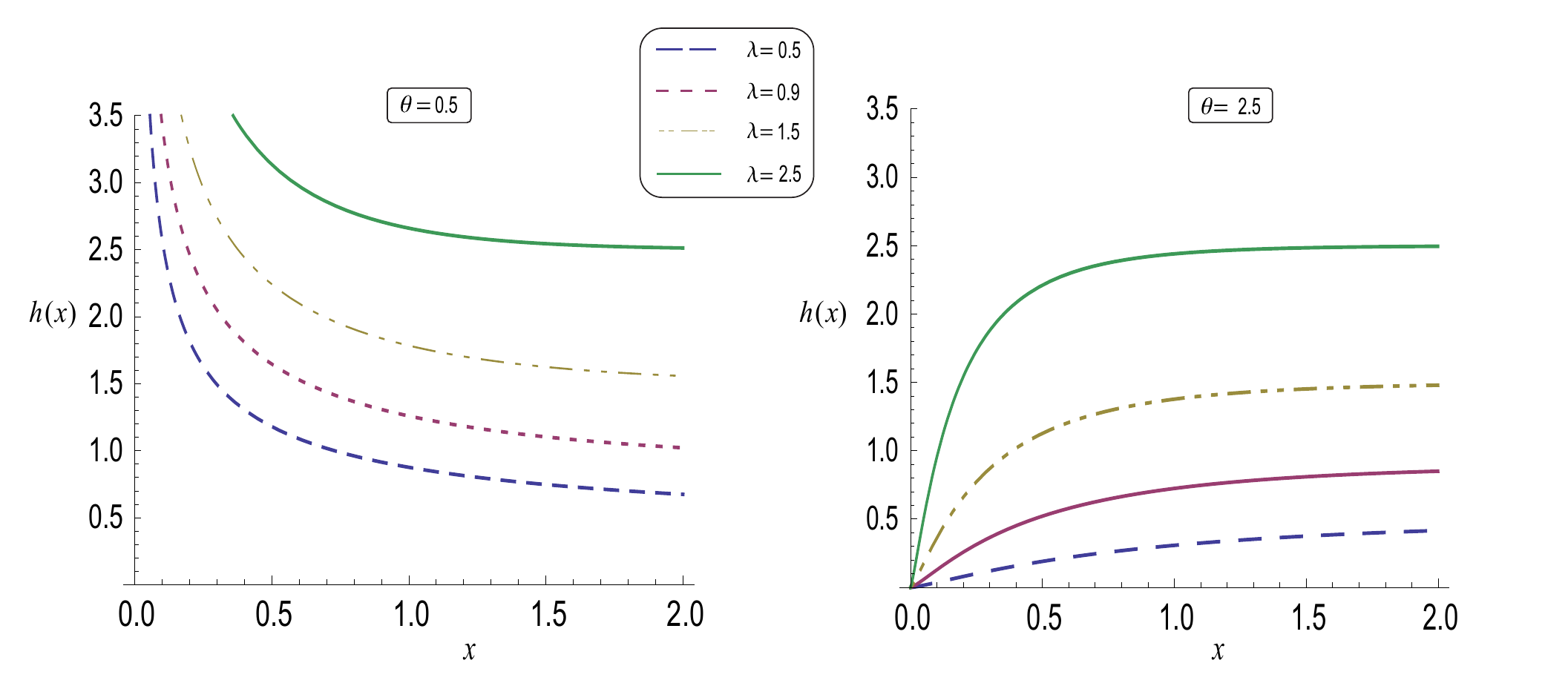}
\caption{Hazard function for various values of $\lambda$ and $\theta$.}
\end{figure}

\section{The Quantile Function of L-E distribution}
The cdf, $F_X(x)=1-\bar{F}(x)$, can be obtained by using eq.(\ref{4}). Further, it can be noted that $F_X$ is continuous and strictly increasing so the quantile function of $ X $ is $Q_X(\gamma)=F^{-1}_X(\gamma)$, $0<\gamma<1$. In the following theorem, we give an explicit expression for $Q_X$ in terms of the Lambert $ W $ function. For more details on Lambert $W$ function we refer the reader to  Jodr\'a \cite{9}.

\noindent \textbf{Theorem 2:} For any $\theta, \lambda >0$, the quantile function of the L-E distribution $ X $ is 
\begin{equation}  \label{8}
x_{\gamma}=F^{-1}(\gamma)=-\frac{1}{\lambda }\log  \left(1-\left(-\frac{W_{-1} \left(-e^{-\theta -1} (\theta +1) \gamma\right)}{(\theta +1) \gamma}\right)^{-1/\theta }\right)
\end{equation}
where $W_{-1}$ denotes the negative branch of the Lambert W function.\\ 
\\
\textit{Proof:} By assuming $p=1-e^{-\lambda x}$, the cdf can be written as 
\begin{equation*}
F_X(x)=\frac{p^\theta(1+\theta-\theta\log(p))}{1+\theta}
\end{equation*}
for fixed $\theta, \lambda >0 $ and $\gamma \in (0,1)$, the $\gamma^{th}$ quantile function is obtained by solving $F_X(x)=\gamma$. By re-arranging the above, we obtain
\begin{equation*}
\log(p^{-\theta})-\gamma p^{-\theta}(1+\theta)=-(1+\theta)
\end{equation*}
taking exponential and multiplying $-(1+\theta)\gamma$  on both sides, we get\\
\begin{equation}  \label{9}
-p^{-\theta}(1+\theta)\gamma e^{-p^{-\theta}(1+\theta)\gamma}=-(1+\theta)\gamma e^{-(1+\theta)}
\end{equation}
By using definition of Lambert-W function ($W(z)e^{W(z)}=z$, where $ z $ is a complex number), we see that $-p^{-\theta}(1+\theta)\gamma$  is the Lambert $ W $ function of the real argument $-(1+\theta)\gamma e^{-(1+\theta)}$. Thus, we have
\begin{equation}  \label{10}
W\left(-(1+\theta)\gamma e^{-(1+\theta)}\right)=-p^{-\theta}(1+\theta)\gamma
\end{equation}
Moreover, for any $\theta, \lambda >0$ it is immediate that  $p^{-\theta}(1+\theta)\gamma > 1 $, and it can also be checked that $(1+\theta)\gamma e^{-(1+\theta)} \in (-1/e,0)$ since $\gamma \in (0,1)$. Therefore, by taking into account the properties of the negative branch of the Lambert W function, we have 
\begin{equation*}
W_{-1}\left(-(1+\theta)\gamma e^{-(1+\theta)}\right)=-p^{-\theta}(1+\theta)\gamma
\end{equation*}
Also by substituting $p=1-e^{-\lambda x}$ in cdf and solving it for $x$, we get
\begin{equation}  \label{11}
x_{\gamma}=F^{-1}(\gamma)=-\frac{1}{\lambda }\log  \left(1-\left(-\frac{W_{-1} \left(-e^{-\theta -1} (\theta +1) \gamma\right)}{(\theta +1) \gamma}\right)^{-1/\theta }\right)
\end{equation} 
\begin{flushright}
$\Box$
\end{flushright}
Further the first three quantiles we obtained by substituting $\gamma =\frac{1}{4},\frac{1}{2},\frac{3}{4} $ in equation (11).

\begin{equation}   \label{12}
\begin{aligned}
Q_1&=F^{-1}(1/4)= - \frac{1}{\lambda}\log\left(1-\left(-\frac{4 W_{-1}\left(\frac{-e^{-\theta-1}(1+\theta)}{4} \right)}{(1+\theta)} \right)^{-\frac{1}{\theta}} \right) \\
\text{Median} (M_d)=Q_2&=F^{-1}(1/2)= -\frac{1}{\lambda}\log\left(1-\left(-\frac{2 W_{-1}\left(\frac{-e^{-\theta-1}(1+\theta)}{2} \right)}{(1+\theta)} \right)^{-\frac{1}{\theta}} \right) \\
Q_3&=F^{-1}(3/4)= -\frac{1}{\lambda}\log\left(1-\left(-\frac{4 W_{-1}\left(\frac{-3e^{-\theta-1}(1+\theta)}{4} \right)}{3(1+\theta)} \right)^{-\frac{1}{\theta}} \right)
\end{aligned}
\end{equation}

\section{Moments}  \label{14}
The moment generating function of the random variable $X$ follow L-E distribution is given as\\
\begin{align} 
\nonumber\mathds{M}_X(t)= E(e^{tX})=&\int_0^{\infty } \frac{\theta ^2 \lambda  e^{-x(\lambda -t)} \left(1-e^{-\lambda  x}\right)^{\theta -1} \left(1-\log \left(1-e^{-\lambda  x}\right)\right)}{\theta +1} \, dx \\
=&\frac{\theta  \Gamma(\theta +1) \Gamma \left(1-\frac{t}{\lambda }\right) \left(-\psi ^{(0)}(\theta )+\psi ^{(0)}\left(-\frac{t}{\lambda }+\theta +1\right)+1\right)}{(\theta +1) \Gamma \left(-\frac{t}{\lambda }+\theta +1\right)}
\end{align}
where, $\psi^{(n)}(z)=\frac{d^n \psi(z)}{dz^n}$ and $\psi(z)=\frac{\Gamma'(z)}{\Gamma(z)}$ known as digamma function.\\
Hence the first and second raw moments can be obtained by $\left(\frac{d\mathds{M}_X(t)}{dx}\right)_{t=0}$ and $\left(\frac{d^2\mathds{M}_X(t)}{dx^2}\right)_{t=0}$ respectively. \\
\begin{align}  \label{15}
\mathds{E}(X)=&\frac{\gamma \theta +\frac{1}{\theta }+\theta  (-\psi ^{(1)}(\theta +1))+(\theta +1) \psi ^{(0)}(\theta )+\gamma +1}{(\theta +1) \lambda } \\
\nonumber \\
\mathds{E}(X^2)=&\frac{{6 \gamma ^2 \theta +\pi ^2 \theta +6 \theta  \psi ^{(0)}(\theta +1)^3+6 (2 \gamma  \theta +\theta ) \psi ^{(0)}(\theta +1)^2-6 \theta  \psi ^{(0)}(\theta ) \psi ^{(0)}(\theta +1)^2-12 \gamma  \theta  \psi ^{(0)}(\theta )^2+ \atop 12 \gamma  (\theta -1) \psi ^{(0)}(\theta )-6 (2 \gamma  \theta +\theta +3) \psi ^{(1)}(\theta +1)-12 \theta  \psi ^{(0)}(\theta ) \psi ^{(1)}(\theta +1)+6 \theta  \psi ^{(2)}(\theta +1)+6 \gamma ^2+\pi ^2+12 \gamma } }{6 (\theta +1) \lambda ^2}
\end{align}
where $\gamma$ is Eulergamma constant =0.577216. \\

\noindent Table 1 displays the mode, mean and median for L-E distribution for different choices of parameter $\lambda$ and $\theta$. It can be observed from the table that all the  three measures of central tendency  decrease with increase in $\lambda$ and increase with an increase in $\theta$. Also for any choice of $\lambda$ and $\theta$ it is observed that Mean $ > $ Median $ > $ Mode , which is an indication of positive skewness.

\begin{table}[htbp]
  \centering
  \caption{Mode, Mean and Median for various value of parameter}
    \begin{tabular}{rcccccccc}

          & \multicolumn{8}{c}{} \\ \hline
 &   $\theta$ $ \downarrow $   &   $\lambda \rightarrow$    & 0.1   & 0.5   & 1     & 1.5   & 2     & 2.5 \\ \hline
    \multicolumn{1}{c}{} & 1.1   & Mode  & \multicolumn{1}{c}{0.001219} & \multicolumn{1}{c}{0.000244} & \multicolumn{1}{c}{0.000122} & \multicolumn{1}{c}{0.000081} & \multicolumn{1}{c}{0.000061} & \multicolumn{1}{c}{0.000049} \\
    \multicolumn{1}{c}{} &       & Mean  & \multicolumn{1}{c}{7.446760} & \multicolumn{1}{c}{1.489350} & \multicolumn{1}{c}{0.744676} & \multicolumn{1}{c}{0.496451} & \multicolumn{1}{c}{0.372330} & \multicolumn{1}{c}{0.297871} \\
    \multicolumn{1}{c}{} &       & Median  & \multicolumn{1}{c}{4.478034} & \multicolumn{1}{c}{0.895607} & \multicolumn{1}{c}{0.447803} & \multicolumn{1}{c}{0.298536} & \multicolumn{1}{c}{0.223902} & \multicolumn{1}{c}{0.179121} \\ \hline
    \multicolumn{1}{c}{} & 1.5   & Mode  & \multicolumn{1}{c}{1.508590} & \multicolumn{1}{c}{0.301719} & \multicolumn{1}{c}{0.150850} & \multicolumn{1}{c}{0.100573} & \multicolumn{1}{c}{0.075429} & \multicolumn{1}{c}{0.060344} \\
    \multicolumn{1}{c}{} &       & Mean  & \multicolumn{1}{c}{9.861580} & \multicolumn{1}{c}{1.972300} & \multicolumn{1}{c}{0.986158} & \multicolumn{1}{c}{0.657438} & \multicolumn{1}{c}{0.493079} & \multicolumn{1}{c}{0.394463} \\
    \multicolumn{1}{c}{} &       & Median  & \multicolumn{1}{c}{6.920488} & \multicolumn{1}{c}{1.384098} & \multicolumn{1}{c}{0.692048} & \multicolumn{1}{c}{0.461366} & \multicolumn{1}{c}{0.346024} & \multicolumn{1}{c}{0.276820} \\ \hline
    \multicolumn{1}{c}{} & 2     & Mode  & \multicolumn{1}{c}{4.174510} & \multicolumn{1}{c}{0.834902} & \multicolumn{1}{c}{0.417451} & \multicolumn{1}{c}{0.278300} & \multicolumn{1}{c}{0.208725} & \multicolumn{1}{c}{0.166980} \\
    \multicolumn{1}{c}{} &       & Mean  & \multicolumn{1}{c}{12.367100} & \multicolumn{1}{c}{2.473421} & \multicolumn{1}{c}{1.236710} & \multicolumn{1}{c}{0.824470} & \multicolumn{1}{c}{0.618350} & \multicolumn{1}{c}{0.494680} \\
    \multicolumn{1}{c}{} &       & Median  & \multicolumn{1}{c}{9.528006} & \multicolumn{1}{c}{1.905601} & \multicolumn{1}{c}{0.952801} & \multicolumn{1}{c}{0.635200} & \multicolumn{1}{c}{0.476400} & \multicolumn{1}{c}{0.381120} \\ \hline
    \multicolumn{1}{c}{} & 2.5   & Mode  & \multicolumn{1}{c}{6.540190} & \multicolumn{1}{c}{1.308040} & \multicolumn{1}{c}{0.654019} & \multicolumn{1}{c}{0.436013} & \multicolumn{1}{c}{0.327010} & \multicolumn{1}{c}{0.261608} \\
    \multicolumn{1}{c}{} &       & Mean  & \multicolumn{1}{c}{14.444000} & \multicolumn{1}{c}{2.888800} & \multicolumn{1}{c}{1.444400} & \multicolumn{1}{c}{0.962930} & \multicolumn{1}{c}{0.722200} & \multicolumn{1}{c}{0.577761} \\
    \multicolumn{1}{c}{} &       & Median  & \multicolumn{1}{c}{11.704978} & \multicolumn{1}{c}{2.340996} & \multicolumn{1}{c}{1.170498} & \multicolumn{1}{c}{0.780332} & \multicolumn{1}{c}{0.585249} & \multicolumn{1}{c}{0.468199} \\ \hline
    \multicolumn{1}{c}{} & 3     & Mode  & \multicolumn{1}{c}{8.569080} & \multicolumn{1}{c}{1.713820} & \multicolumn{1}{c}{0.856908} & \multicolumn{1}{c}{0.571272} & \multicolumn{1}{c}{0.428454} & \multicolumn{1}{c}{0.342763} \\
    \multicolumn{1}{c}{} &       & Mean  & \multicolumn{1}{c}{16.204700} & \multicolumn{1}{c}{3.240930} & \multicolumn{1}{c}{1.620470} & \multicolumn{1}{c}{1.080310} & \multicolumn{1}{c}{0.810233} & \multicolumn{1}{c}{0.648186} \\
    \multicolumn{1}{c}{} &       & Median  & \multicolumn{1}{c}{13.549240} & \multicolumn{1}{c}{2.709848} & \multicolumn{1}{c}{1.354924} & \multicolumn{1}{c}{0.903283} & \multicolumn{1}{c}{0.677462} & \multicolumn{1}{c}{0.541970} \\ \hline
    \multicolumn{1}{c}{} & 3.5   & Mode  & \multicolumn{1}{c}{10.317200} & \multicolumn{1}{c}{2.063400} & \multicolumn{1}{c}{1.031720} & \multicolumn{1}{c}{0.687811} & \multicolumn{1}{c}{0.515858} & \multicolumn{1}{c}{0.412687} \\
    \multicolumn{1}{c}{} &       & Mean  & \multicolumn{1}{c}{17.726300} & \multicolumn{1}{c}{3.545200} & \multicolumn{1}{c}{1.772630} & \multicolumn{1}{c}{1.181760} & \multicolumn{1}{c}{0.886310} & \multicolumn{1}{c}{0.709053} \\
    \multicolumn{1}{c}{} &       & Median  & \multicolumn{1}{c}{15.138649} & \multicolumn{1}{c}{3.027730} & \multicolumn{1}{c}{1.513865} & \multicolumn{1}{c}{1.009243} & \multicolumn{1}{c}{0.756932} & \multicolumn{1}{c}{0.605546} \\ \hline
    \multicolumn{1}{c}{} & 4     & Mode  & \multicolumn{1}{c}{11.840500} & \multicolumn{1}{c}{2.368100} & \multicolumn{1}{c}{1.184050} & \multicolumn{1}{c}{0.789366} & \multicolumn{1}{c}{0.592025} & \multicolumn{1}{c}{0.473620} \\
    \multicolumn{1}{c}{} &       & Mean  & \multicolumn{1}{c}{19.062700} & \multicolumn{1}{c}{3.812500} & \multicolumn{1}{c}{1.906270} & \multicolumn{1}{c}{1.270850} & \multicolumn{1}{c}{0.953130} & \multicolumn{1}{c}{0.762510} \\
    \multicolumn{1}{c}{} &       & Median  & \multicolumn{1}{c}{16.529903} & \multicolumn{1}{c}{3.305981} & \multicolumn{1}{c}{1.652990} & \multicolumn{1}{c}{1.101994} & \multicolumn{1}{c}{0.826495} & \multicolumn{1}{c}{0.661196} \\ \hline
    \end{tabular}%
  \label{tab:addlabel}%
\end{table}%

\section{Limiting Distribution of Sample Minima and Maxima}
We can derive the asymptotic distribution of the sample minimum $X_{1:n}$ by using theorem 8.3.6 of Arnold t. al.\cite{2}, it follows that the asymptotic distribution of $X_{1:n}$ is Weibull type with shape parameter $\theta>0$ if 
\begin{equation*}
\lim \limits_{t \to 0^+}\frac{F(tx)}{F(t)}=x^\theta
\end{equation*}
for all $x > 0$. Then, by using L ’H\'{o}pital’s rule, it follows that
\begin{equation*}
\lim \limits_{t \to 0^+}\frac{F(tx)}{F(t)}=x \lim \limits_{t \to 0^+}\frac{f(tx)}{f(t)}= x \lim \limits_{t \to 0^+} \frac{(1-e^{-\lambda t x})^{\theta-1}(1-\log(1-e^{-\lambda t x}))}{(1-e^{-\lambda t })^{\theta-1}(1-\log(1-e^{-\lambda t}))}=x^{\theta}
\end{equation*}
Since
\begin{align*}
\lim \limits_{t \to 0^+} \frac{(1-e^{-\lambda t x})^{\theta-1}}{(1-e^{-\lambda t })^{\theta-1}}=x^{\theta-1}
\end{align*}
and 
\begin{align*}
\lim \limits_{t \to 0^+} \frac{(1-\log(1-e^{-\lambda t x}))}{(1-\log(1-e^{-\lambda t}))}=1
\end{align*}
Hence, we obtain that the asymptotic distribution of the sample minima $X_{1:n}$ is of the Weibull type with
shape parameter $\theta$. \\
Further, it can be seen that 
\begin{equation*}
\lim \limits_{t \to \infty}\frac{1-F(t+x)}{1-F(t)}=\lim \limits_{t \to \infty} \frac{(1+\theta)-(1-e^{-\lambda(x+t)})^\theta(1+\theta-\theta\log(1-e^{-\lambda(x+t)}))}{(1+\theta)-(1-e^{-\lambda t})^\theta(1+\theta-\theta\log(1-e^{-\lambda t}))}
\end{equation*}
by using L-H\'{o}pital’s rule, 
\begin{equation*}
\lim \limits_{t \to \infty}\frac{e^{-\lambda x} \left(1-e^{-\lambda(t+x)}\right)^{\theta -1} \left(1-\log \left(1-e^{-\lambda(t+x)}\right)\right)}{\left(1-e^{-\lambda t}\right)^{\theta-1} \left(1-\log \left(1-e^{-\lambda t)}\right)\right)}=e^{-\lambda \theta x}
\end{equation*}
Since
\begin{equation*}
\lim \limits_{t \to \infty} \frac{\left(1-e^{-\lambda(t+x))}\right)^{\theta -1}}{\left(1-e^{-\lambda t}\right)^{\theta-1}} =e^{-\lambda x(\theta-1)}
\end{equation*}
and
\begin{equation*}
\lim \limits_{t \to \infty} \frac{1-\log \left(1-e^{\lambda  (-(t+x))}\right)}{1-\log \left(1-e^{\lambda  (-t)}\right)} = 1
\end{equation*}
Hence, it follows from Theorem 1.6.2 in Leadbetter et al. (1983) that there must be norming constants $a_n$,$ b_n, c_n>0$ and $d_n$ such that
\begin{equation}  \label{16}
\text{Pr}\lbrace a_n(M_n-b_n)\le x \rbrace \rightarrow e^{-e^{-\theta \lambda x}}
\end{equation}
and
\begin{equation}  \label{17}
\text{Pr}\lbrace c_n(m_n-d_n)\le x \rbrace \rightarrow 1-e^{-x^\theta} 
\end{equation}
as $n \rightarrow \infty$. By following Corollary 1.6.3 in Leadbetter et al. (1983), we can determine the form of the norming constants. As an illustration, one can see that $a_n = \theta$ and $b_n=F^{-1}(1-1/n)$, where $ F^{-1}(.) $ denotes the inverse function of $ F(.) $.

\section{Entropy}
In many field of science such as communication, physics and probability, entropy is an important concept to measure the amount of uncertainty associated with a random variable $X$. Several entropy measures and information indices are available but among them the most popular entropy measure called  R\'enyi entropy is defined as 
\begin{equation} \label{18}
\mathfrak{J}(\zeta)=\frac{1}{1-\zeta} \log \left( \, \int\limits_{\mathbb{R}^+}f^\zeta(x) dx \right),  \quad \text{for} \quad \zeta >1 \quad \text{and} \quad \zeta \ne 1
\end{equation}
In our case 
\begin{equation*}
\int\limits_{\mathbb{R}^+} f^\zeta dx= \int\limits_{\mathbb{R}^+} \left(\frac{\theta^2 \lambda}{\theta+1}\right)^{\zeta}e^{-x \lambda \zeta} \left(1-e^{-x\lambda}\right)^{\zeta(\theta -1)} \left(1-\log \left(1-e^{-x\lambda}\right)\right)^\zeta dx
\end{equation*}
substituting $x=-\frac{1}{\lambda}\log(1-e^{-u})$ and using power series expansion $(1-z)^\alpha=\sum\limits_{j=0}^{\infty}(-1)^j \binom \alpha j z^j$, the above expression reduces to 
\begin{align} \label{19}
\nonumber \int\limits_{\mathbb{R}^+} f^\zeta dx=& \frac{\theta ^{2 \zeta } \lambda ^{\zeta -1}}{(\theta +1)^{\zeta }}\int_0^{\infty } (u+1)^{\zeta} \left(1-e^{-u}\right)^{\zeta -1} e^{u (-(\zeta (\theta-1)+1))} \, du \\ \nonumber
=& \frac{\theta ^{2 \zeta } \lambda ^{\zeta-1}}{(\theta +1)^{\zeta}}\int _0^{\infty }\sum _{j=0}^{\infty } (-1)^j \binom{\zeta-1}{j} e^{u(-(\zeta(\theta -1)+j+1))} (u+1)^{\zeta} du  \\ \nonumber
=&\frac{\theta ^{2 \zeta} \lambda ^{\zeta-1}}{(\theta +1)^{\zeta}} \sum _{j=0}^{\infty } (-1)^j \binom{\zeta-1}{j} \int_0^{\infty } (u+1)^{\zeta} e^{u(-(\zeta(\theta -1)+j+1))} \, du \\
=&\frac{\theta ^{2 \zeta } \lambda ^{\zeta -1}}{(\theta +1)^{\zeta }} \sum_{j=0}^{\infty} (-1)^j e^{\zeta(\theta -1)+j+1} E_{-\zeta }\left[(j+(\theta -1) \zeta +1)\right]
\end{align}
where $E_n(z)=\int_1^{\infty }e^{-zt} t^{-n} dt$ known as exponential integral function. For more details \\
see http://functions.wolfram.com/06.34.02.0001.01. \\
Thus according to (18) the R\'enyi entropy of L-E$(\theta,\lambda)$ distribution is given by 
\begin{equation} \label{20}
\mathfrak{J}(\zeta)=\frac{1}{1-\zeta}\log\left(\frac{\theta ^{2\zeta} \lambda^{\zeta-1}}{(\theta +1)^{\zeta }} \sum_{j=0}^{\infty} (-1)^j e^{\zeta(\theta -1)+j+1} E_{-\zeta}\left[(j+(\theta -1) \zeta +1)\right]\right)
\end{equation}
Moreover, the Shannon entropy is defined by $E[−\log(f(x))]$. This is a special case derived from $\lim\limits_{\zeta \rightarrow 1}\mathfrak{J}(\zeta)$

\section{Maximum Likelihood function}
In this section we shall discuss the point and interval estimation on the parameters that index the L-E$ (\theta,\lambda)$. Let the log-likelihood function $L=l(\Theta)$ of single observation(say $ x_i $) for the vector of parameter $\Theta=(\theta, \lambda)^\top$ can be written as
\begin{equation} \label{21}
\nonumber l_n(x,\Theta)=2 \log(\theta )+ \log (\lambda)-\lambda x_i -\log (\theta +1)+(\theta-1) \log (1-e^{-\lambda x_i })
+\log (1-\log (1-e^{-\lambda  x_i})), \quad x_i>0
\end{equation}
The associated score function is given by $U_n(\Theta)= \left(\frac{\partial l_n}{\partial \theta}, \frac{\partial l_n}{\partial \lambda} \right)^\top$, where 
\begin{align} \label{22}
\frac{\partial l_n}{\partial \theta}&=\frac{2}{\theta}-\frac{1}{\theta +1}+ \log \left(1-e^{-\lambda  x_i}\right) \\
\frac{\partial l_n}{\partial \lambda}&=\frac{1}{\lambda }-x_i+(\theta-1)\frac{x_i e^{-\lambda  x_i}}{1-e^{-\lambda x_i }}- \frac{x_i e^{-\lambda x}}{\left(1-e^{-\lambda  x_i}\right)(1- \log \left(1-e^{-\lambda  x_i}\right))}
\end{align}
As we know th xpctd value of score function equals zero, i.e. $\mathtt{E}(U(\Theta)=0)$, which implies \\
$\mathds{E}\left( \log \left(1-e^{-\lambda  x}\right)\right)= \frac{1}{\theta +1}-\frac{2}{\theta } $ \\

The total log-likelihood of the random sample $x=\left(x_1,\cdots, x_n\right)^\top$ of size $ n $ from $ X $ is given by $l_n= \sum\limits_{1}^{n}l^{(i)}$ and th total score function is given by $U_n=\sum\limits_{i=1}^{n}U^{(i)}$, where $l^{(i)}$ is the log-likelihood of $i^{th}$ observation. The maximum likelihood estimator $\hat{\Theta}$ of $\Theta$ is obtained by solving equation(21) and (22) numerically or this can also be obtained easily by using nlm() function in R. Moreover the Fisher information matrix is given by 
\begin{equation} \label{23}
K=K_n(\Theta)= n \left[ \begin{matrix}
\kappa_{\theta,\theta}&  \kappa_{\theta,\lambda} \\ 
\kappa_{\lambda,\theta}  & \kappa_{\lambda,\lambda}
\end{matrix} \right]
\end{equation}
where
\begin{equation} \label{24} 
\begin{aligned} 
\kappa_{\theta,\theta}&= \frac{2}{(\theta +1)^2}-\frac{2}{\theta ^2} \\
\kappa_{\lambda,\theta}&=  \mathds{E} \left( \frac{X e^{-\lambda X}}{1-e^{-\lambda X}}\right) \\
\kappa_{\lambda,\lambda}&= \frac{1}{\lambda^2}+(\theta-1)\mathtt{E}\left(\frac{X^2 e^{-\lambda  X}}{\left(1-e^{-\lambda X}\right)^2}\right)-\mathds{E}\left(\frac{e^{- \lambda X} X^2 \left(\log \left(1-e^{-\lambda  X}\right)+e^{- \lambda X}\right)}{\left(1-e^{-\lambda X}\right)^2 \left(\log\left(1-e^{-\lambda  X}\right)\right)^2}\right)
\end{aligned}
\end{equation}
The above expressions depend on some expectations which easily computed using numerical integration. Under the usual regularity conditions, the asymptotic distribution of
\begin{equation}
\sqrt{n}\left( \hat{\Theta}-\Theta\right) \, \,  \text{is} \, \,  N_2(0,K(\Theta)^{-1})
\end{equation}
where $\lim\limits_{n \rightarrow \infty}=K_n(\Theta)^{-1}=K(\Theta)^{-1}$. The asymptotic multivariate normal $N_2(0,K(\Theta)^{-1})$  distribution of $\hat{\Theta}$ can be usd to construct approximate confidence  intervals. An asymptotic confidence interval with significance level $ \alpha $ for each
parameter $ \theta $ and $ \lambda $ is
\begin{equation}
\begin{aligned}
\text{ACI}\left(\theta,100(1-\alpha)\% \right)=&(\hat{\theta}-z_{\alpha/2}\sqrt{\kappa_{(\theta,\theta)}},\hat{\theta}+z_{\alpha/2}\sqrt{\kappa(_\theta,\theta)}) \\
\text{ACI}\left(\lambda,100(1-\alpha)\%
\right)=&(\hat{\lambda}-z_{\alpha/2}\sqrt{\kappa_{(\lambda,\lambda)}},\hat{\lambda}+z_{\alpha/2}\sqrt{\kappa(_\lambda,\lambda)})
\end{aligned}
\end{equation}
where $z_{1-\alpha/2}$ denotes $1-\alpha/2$ quantile of standard normal random variable.

\section{Simulation}
In this section, we investigate the behavior of the ML estimators for a finite sample size ($ n $). Simulation study based on different L-E$(\theta,\lambda)$ distribution is carried out.  The random variable are generated by using cdf technique  presented in section 4 from  L-E$(\theta,\lambda)$ are generated. A simulation study consisting of following steps is being carried out for each triplet $(\theta,\lambda,n)$, where $\theta= 0.5, 1, 2, \lambda= 0.5, 1, 2 ,3$ and $n= 20, 50, 75, 100$.
\begin{enumerate}
\item Choose the initial values  of $ \theta_\circ,\lambda_\circ $ for the corresponding elements of the parameter vector $\Theta=(\theta,\lambda)$, to specify L-E$(\theta,\lambda)$ distribution;
\item choose sample size $ n $;
\item generate $ N $ independent samples of size $ n $ from L-E$(\theta,\lambda)$;
\item compute the ML estimate $\hat{\Theta_n}$ of $\Theta_\circ$ for each of the $ N $ samples;
\item compute the mean of the obtained estimators over all $ N $ samples, the average bias $= \frac{1}{N} \sum\limits_{i=1}^{N}(\Theta_i-\Theta_\circ) $ and the average mean square error $MSE$ $(\Theta)= \frac{1}{N} \sum\limits_{i=1}^{N}(\Theta_i-\Theta_\circ)^2$, of simulated estimates.
\end{enumerate}

\begin{table}[htbp]
  \centering
  \caption{Average bias of the simulated estimates.}
    \begin{tabular}{ccccccccccccc} \hline
          & $ n $     & \multicolumn{2}{c}{$\lambda=0.5$} & & \multicolumn{2}{c}{$\lambda=1$} & & \multicolumn{2}{c}{$\lambda=2$} & &\multicolumn{2}{c}{$\lambda=3$} \\ 
          \cline{3-4} \cline{6-7} \cline{9-10} \cline{12-13}
          &       & \multicolumn{1}{c}{bais $(\theta)$} & \multicolumn{1}{c}{bais $(\lambda)$} & & \multicolumn{1}{c}{bais $(\theta)$} & \multicolumn{1}{c}{bais $(\lambda)$} & & \multicolumn{1}{c}{bais $(\theta)$} & \multicolumn{1}{c}{bais $(\lambda)$} & & \multicolumn{1}{c}{bais $(\theta)$} & \multicolumn{1}{c}{bais $(\lambda)$} \\ \hline
    \multirow{4}{*}{$\theta=0.5$} & 20 & 0.0448 & 0.1836 & & 0.0476 & 0.4173 & & 0.0519 & 0.8777 & & 0.0534 & 1.1682 \\
          & 50    & 0.0187 & 0.0673 & &0.0196 & 0.1596 & &0.0164 & 0.2694 & &0.0148 & 0.3427 \\
          & 75    & 0.0092 & 0.0403 & &0.0131 & 0.0792 & &0.0136 & 0.2166 & &0.0120 & 0.2608 \\
          & 100   & 0.0067 & 0.0267 & &0.0091 & 0.0728 & &0.0074 & 0.1318 & &0.0095 & 0.1988 \\ \hline
    \multirow{4}{*}{$\theta=1$}  & 20    & 0.1200 & 0.1012 & &0.1073 & 0.1873 & &0.1158 & 0.3884 & &0.1085 & 0.5350 \\
          & 50    & 0.0478 & 0.0405 & & 0.0412 & 0.0655 & &0.0507 & 0.1723 & &0.0522 & 0.2444 \\
          & 75    & 0.0364 & 0.0268 & & 0.0265 & 0.0370 & &0.0286 & 0.1091 & &0.0285 & 0.1415 \\
          & 100   & 0.0150 & 0.0136 & & 0.0198 & 0.0315 & &0.0183 & 0.0548 & &0.0244 & 0.1266 \\ \hline
   \multirow{4}{*}{$\theta=2$}  & 20    & 0.3599 & 0.0650 & &0.3270 & 0.1268 & &0.3628 & 0.2334 & &0.3401 & 0.3451 \\
          & 50    & 0.1003 & 0.0204 & &0.1100 & 0.0460 & &0.1200 & 0.1021 & &0.1209 & 0.1324 \\
          & 75    & 0.0686 & 0.0136 & &0.0654 & 0.0312 & &0.0803 & 0.0653 & &0.0955 & 0.1173 \\
          & 100   & 0.0562 & 0.0136 & &0.0457 & 0.0234 & &0.0511 & 0.0417 & &0.0588 & 0.0922 \\ \hline
    \end{tabular}%
  \label{tab:addlabel}%
\end{table}%

\begin{table}[htbp]
  \centering
  \caption{Average MSE of the simulated estimates.}
    \begin{tabular}{ccccccccccccc} \hline
          & \multicolumn{1}{c}{$ n $} & \multicolumn{2}{c}{$\lambda=0.5$} & \multicolumn{1}{c}{} & \multicolumn{2}{c}{$\lambda=1$} & \multicolumn{1}{c}{} & \multicolumn{2}{c}{$\lambda=2$} & \multicolumn{1}{c}{} & \multicolumn{2}{c}{$\lambda=3$} \\
			\cline{3-4} \cline{6-7} \cline{9-10} \cline{12-13}
          &       & \multicolumn{1}{c}{MSE$(\theta)$} & \multicolumn{1}{c}{MSE$(\lambda)$} & \multicolumn{1}{c}{} & \multicolumn{1}{c}{MSE$(\theta)$} & \multicolumn{1}{c}{MSE$(\lambda)$} & \multicolumn{1}{c}{} & \multicolumn{1}{c}{MSE$(\theta)$} & \multicolumn{1}{c}{MSE$(\lambda)$} & \multicolumn{1}{c}{} & \multicolumn{1}{c}{MSE$(\theta)$} & \multicolumn{1}{c}{MSE$(\lambda)$} \\ \hline
    \multirow{4}{*}{$\theta=0.5$} & \multicolumn{1}{c}{20} & 0.0164 & 0.1988 &  & 0.0177 & 1.0549 &       & 0.0204 & 5.3140 & & 0.0200 & 8.3846 \\
          & \multicolumn{1}{c}{50} & 0.0051 & 0.0459 &  & 0.0055 & 0.2089 &  & 0.0053 & 0.7757 &  & 0.0051 & 1.5790 \\
          & \multicolumn{1}{c}{75} & 0.0032 & 0.0270 &  & 0.0031 & 0.1047 &  & 0.0033 & 0.4656 &  & 0.0032 & 0.9631 \\
          & \multicolumn{1}{c}{100} & 0.0024 & 0.0161 &  & 0.0022 & 0.0763 &  & 0.0022 & 0.2870 &  & 0.0024 & 0.7081 \\ \hline
    \multirow{4}{*}{$\theta=1$} & \multicolumn{1}{c}{20} & 0.1313 & 0.0802 &  & 0.0998 & 0.2457 &  & 0.1012 & 1.1482 &       & 0.1067 & 2.2693 \\
          & \multicolumn{1}{c}{50} & 0.0314 & 0.0200 &  & 0.0283 & 0.0716 &  & 0.0302 & 0.3081 &  & 0.0316 & 0.7259 \\
          & \multicolumn{1}{c}{75} & 0.0195 & 0.0119 &  & 0.0163 & 0.0431 &  & 0.0178 & 0.1875 &  & 0.0171 & 0.3556 \\
          & \multicolumn{1}{c}{100} & 0.0112 & 0.0078 &  & 0.0128 & 0.0331 &  & 0.0111 & 0.1131 &  & 0.0125 & 0.3082 \\ \hline
    \multirow{4}{*}{$\theta=2$} & \multicolumn{1}{c}{20} & 0.8890 & 0.0344 &       & 0.7999 & 0.1393 &       & 0.9890 & 0.5046 &       & 0.9142 & 1.1439 \\
          & \multicolumn{1}{c}{50} & 0.1733 & 0.0097 &  & 0.1815 & 0.0425 &  & 0.1633 & 0.1597 &  & 0.1857 & 0.3790 \\
          & \multicolumn{1}{c}{75} & 0.1131 & 0.0063 &  & 0.1008 & 0.0244 &  & 0.1075 & 0.1069 &  & 0.1028 & 0.2132 \\
          & \multicolumn{1}{c}{100} & 0.0730 & 0.0046 &  & 0.0712 & 0.0185 &  & 0.0730 & 0.0785 &  & 0.0625 & 0.1586 \\ \hline
    \end{tabular}%
  \label{tab:addlabel}%
\end{table}%

\section{Application to Real Datasets}
In this section, we illustrate, the applicability of L-E Distribution by considering two different datasets used by different researchers. We also fit L-E distribution, Power-Lindley distribution \cite{6} , New Generalized Lindley Distribution \cite{4}, Lindley Distribution, Weibull distribution and Exponential distribution. Namely\\
(i) Power-Lindley distribution (PL$(\alpha,\beta)$):
\begin{equation*}
f_1(x)=\frac{\alpha \beta^2}{1+\beta}(1+x^\alpha) x^{\alpha-1} e^{-\beta x^\alpha},  \quad \quad x,\alpha,\beta >0.
\end{equation*}

\noindent (ii) New Generalized Lindley distribution (NGLD($\alpha,\beta,\theta$)):
\begin{equation*}
f_2(x)=\frac{e^{-\theta x}}{1+\theta}\left(\frac{\theta^{\alpha+1}x^{\alpha-1}}{\Gamma(\alpha)}+\frac{\theta^{\beta}x^{\beta-1}}{\Gamma(\beta)}\right), \quad \quad x,\alpha,\theta,\beta >0
\end{equation*}

\noindent(iii) Lindley Distribution (L$(\theta)$)
\begin{equation*}
f_3(x)=\frac{\theta^2}{1+\theta}(1+x) e^{-\theta x},  \quad \quad x,\alpha,\beta >0.
\end{equation*}

In each of these distributions, the parameters are estimated by using the maximum likelihood method, and for comparison we use negative log-likelihood values ($-LL$), the Akaike information criterion (AIC) and Bayesian information criterion (BIC) which are defined by $-2LL+2q$ and $-2LL+q\log(n)$, respectively, where $q$ is the number of parameters estimated and $n$ is the sample size. Further K-S(Kolmogorov-Smirnov) test statistic defined as $K-S= \sup_x|F_n(x)-F(x)|$, where $F_n(x)=\frac{1}{n} \sum\limits_{i=1}^{n}\textbf{I}_{x_i\le x}$ is empirical distribution function and $F(x)$ is cumulative distribution function is calculated and shown for all the datasets.

\subsection{Illustration 1}
We consider an uncensored data set corresponding to remission times (in months) of a random sample of 128 bladder cancer patients(Lee and Wang\cite{10}) as presented in Appendix A.1.The data sets are presented in appendix A.1 in Table (6). The results for these data are presented in Table 4. We observe that the L-E distribution is a competitive distribution as  compared with other distributions. In fact, based on the values of the AIC, BIC and as well as the value of the K-S test statistic, we observe that the L-E distribution provides the best fit for these data among all the models considered. In Figure 2, we have plotted probability density function and empirical distribution function for all considered distributions for these data.

\begin{table}[htbp]
  \centering
  \caption{The estimates of parameters and goodness-of-fit statistics for Illustration 1.}
    \begin{tabular}{|l|l|c|c|c|c|} \hline
    Model & Parameter & -LL  & AIC   & BIC   & K-S statistic \\ \hline
    L-E    & $ \hat{\lambda} $= 0.0962, $ \hat{\theta} $=1.229 & \multicolumn{1}{c}{401.78} & \multicolumn{1}{|c|}{807.564} & \multicolumn{1}{|c|}{807.780} & \multicolumn{1}{|r|}{0.0454} \\
    PL    & $\hat{\theta} $=0.385, $\hat{\beta}$=0.744 & \multicolumn{1}{|c|}{402.24} & \multicolumn{1}{|c|}{808.474} & \multicolumn{1}{|c|}{808.688} & \multicolumn{1}{|r|}{0.0446} \\
    L     & $\hat{\theta}$=0.196 & \multicolumn{1}{|c|}{419.52} & \multicolumn{1}{|c|}{841.040} & \multicolumn{1}{|c|}{843.892} & \multicolumn{1}{|r|}{0.0740} \\
    \multicolumn{1}{|l|}{NGLD} & $ \hat{\theta} $=0.180, $\hat{\alpha} $=4.679, $ \hat{\beta} $=1.324 & \multicolumn{1}{|c|}{412.75} & \multicolumn{1}{|c|}{831.501} & \multicolumn{1}{|c|}{840.057} & \multicolumn{1}{|r|}{0.1160} \\ \hline
    \end{tabular}%
  \label{tab:addlabel}%
\end{table}%

\begin{figure}
\centering
\includegraphics[width=1\textwidth]{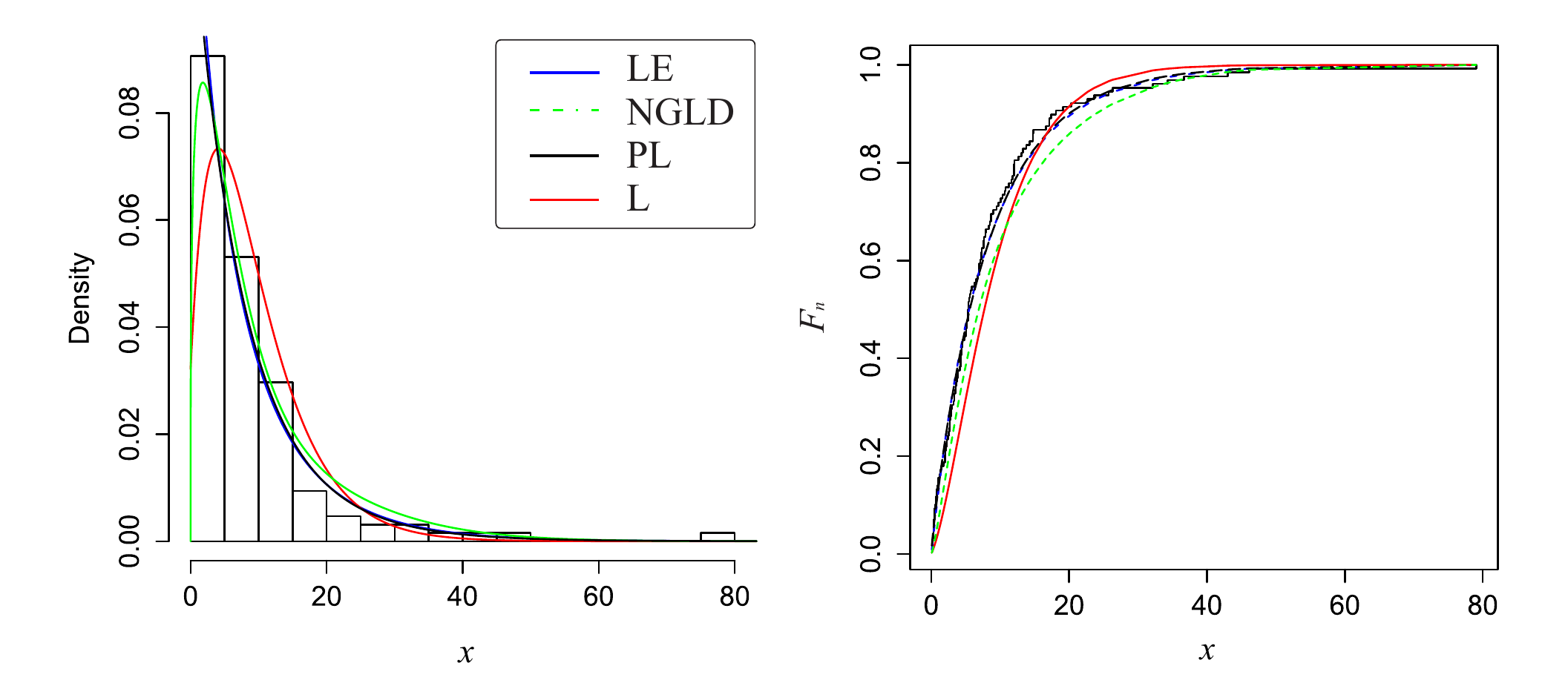}
\caption{PDF plot for various values of $\lambda$ and $\theta$.}
\end{figure}

\subsection{Illustration 2}
As second example, we consider 100 observations on waiting time (in minutes) before the customer service in a bank (see Ghitany et al.\cite{7}). The data sets are presented in appendix A.2 in Table (7). The results for these data are presented in Table 5. From these results we can observe that L-E distribution provide smallest AIC and BIC values as compare to Power lindley, new generalized Lindley distribution, Lindley and exponential  and hence best fits the data among all the models considered. The results are presented in Table 5 and probability density function and empirical distribution function are shown in Figure 3.

\begin{table}[htbp]
  \centering
  \caption{The estimates of parameters and goodness-of-fit statistics for Illustration 2.}
    \begin{tabular}{|l|l|r|r|r|r|} \hline
    Model & Parameter & -LL    & AIC   & BIC   & K-S \\ \hline
    L-E    & $\hat{\theta}$=2.650; $\hat{\lambda}$=0.1520 & 317.005 & 638.01 & 638.1337  & 0.0360 \\
    PL    & $\hat{\theta}$=0.1530;$\hat{\beta}$=1.0832 & 318.319 & 640.64  & 640.64 & 0.0520 \\
    L & $\hat{\theta}$=0.187 & 319.00   & 640.00   & 640.00   & 0.0680 \\
    E & $\hat{\theta}$=0.101 & 329.00   & 660.00   & 660.00   & 0.1624 \\
    NGLD  & $\hat{\theta}$= 0.2033; $\hat{\beta}$=2.008; $\hat{\alpha}$=2.008 & 317.3 & 640.60  & 640.60 & 0.0425 \\ \hline
    \end{tabular}%
  \label{tab:addlabel}%
\end{table}%
\begin{figure}
\centering
\includegraphics[width=1\textwidth]{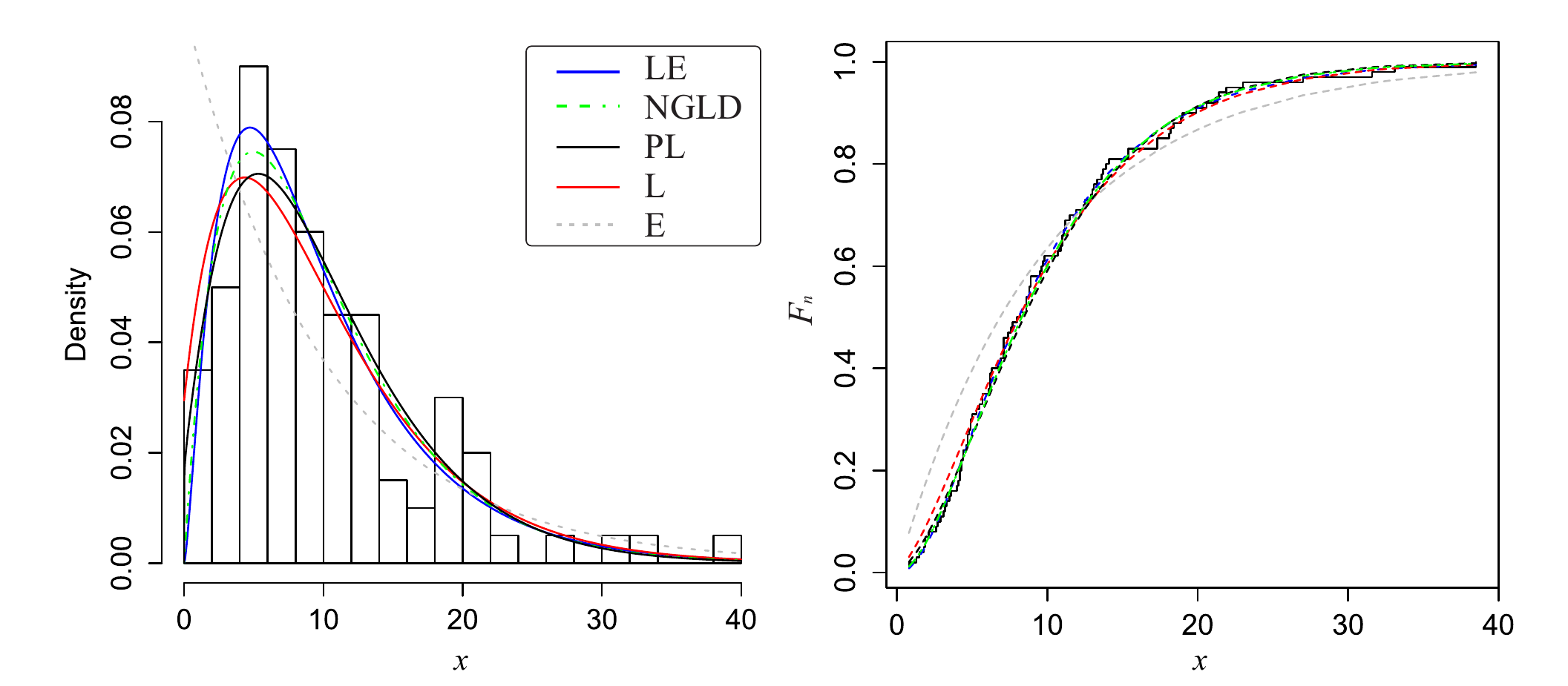}
\caption{PDF plot for various values of $\lambda$ and $\theta$.}
\end{figure}

\section{Estimation of the Stress-Strength Parameter} 
The stress-strength parameter $(R)$ plays an important role in the reliability analysis as it measures the system performance. Moreover, $R$ provides the probability of a system failure,  the system fails whenever the applied stress is greater than its strength i.e. $R=P\left(X>Y\right)$. Here $ X\sim $L-E$(\theta_1,\lambda)$ denotes the strength of a system subject to stress $Y$, and $Y\sim$L-E$(\theta_2,\lambda)$, X and Y are independent of each other. In our case, the stress-strength parameter
R is given by
\begin{equation}
\begin{aligned}
R&=P\left(X>Y\right)=\int\limits_{0}^{\infty} P\left(X >Y | Y=y\right) f_Y(y) dy \\
&=\int\limits_{0}^{\infty}S_X(y) f_Y(y) dy \\
&=1-\int_0^{\infty} \frac{\theta^2_2 \lambda \left(1-e^{-{\lambda y}}\right)^{\theta_1+\theta_2-1} e^{-\lambda y}\left(1-\log \left(1-e^{-\lambda y}\right)\right)\left(1+\theta_1 (1-\log \left(1-e^{-{\lambda y}})\right)\right)}{\left(1+\theta _1\right) \left(1+\theta _2\right)} dy \\
=&1-\frac{\theta _2^2 \left(\theta _1^3+\left(2 \theta _2+3\right) \theta _1^2+\left(\theta _2+1\right) \left(\theta _2+3\right) \theta _1+\theta _2^2+\theta _2\right)}{\left(\theta _1+1\right) \left(\theta _2+1\right) \left(\theta _1+\theta _2\right){}^3}
\end{aligned}
\end{equation}
Remarks: \\
(i) R  is independent of $ \lambda $ \\
(ii) When $\theta_1=\theta_2$, R=0.5. This is intuitive that X and Y are i.i.d. and there is an equal chance that X is bigger than Y. \\

\noindent Since R in equation (27) is a function of stress-strength parameters $\theta_1$ and $\theta_2$ we need to obtain the maximum likelihood estimators (MLEs) of $\theta_1$ and $\theta_2$ to compute the MLE of R under invariance property of the MLE. Suppose that $ X_1,X_2,\cdots,X_n $ and $Y_1,Y_2,\cdots,Y_m$ are independent random samples from L-E $(\theta_1, \lambda)$ and L-E $(\theta_2,\lambda )$ respectively. Thus, the likelihood function based on the observed sample is given by \\

\begin{align*}
L(\theta_1,\theta_2)=\frac{\theta_2^{2m} \theta_1^{2n} \lambda^{m+n}}{\left(\theta_1+1\right) \left(\theta_2+1\right)} &\prod_{i=1}^n e^{-\lambda x_i} \left(1-e^{-\lambda x_i}\right)^{\theta_1-1} \left(1-\log \left(1-e^{-\lambda x_i}\right)\right)  \\ & \cdotp  \prod_{j=1}^m e^{-\lambda y_j} \left(1-e^{-\lambda y_j}\right)^{\theta_2-1} \left(1-\log \left(1-e^{-\lambda y_j}\right)\right)
\end{align*}
\noindent The log - likelihood function is given by
\begin{align*}
\log L(\theta _1,\theta _2)=&2 n \log \left(\theta _1\right)+2 m \log \left(\theta _2\right)+ (m+n)\log(\lambda)-n\log\left(\theta_1+1\right)-m \log\left(\theta_2+1\right)\\ &-\left(\theta_1-1\right) s_1-\lambda \sum_{i=1}^n x_i+\sum_{i=1}^n \log (1-\log(1-e^{-\lambda x_i}))-\left(\theta _2-1\right)s_2 -\lambda \sum_{j=1}^m y_j\\ & +\sum _{j=1}^m \log (1-\log(1-e^{-\lambda y_j}))
\end{align*}
\noindent where $s_1=- \sum\limits_{i=1}^{n} \log \left(1-e^{-\lambda  x_i}\right)$ and $ s_2=-\sum\limits_{j=1}^{m}\log \left(1-e^{-\lambda  y_j}\right)$. \\
\noindent The MLE of $\theta_1$ and $theta_2$, say $\hat{\theta_1}$ and $\hat{\theta_2}$ respectively, can be obtained as the solutions of the following equations
\begin{equation}
\begin{aligned}
\frac{\partial \text{logL}}{\partial \theta _1}&=\frac{2 n}{\theta _1}-\frac{n}{\theta _1+1}-s_1 \\
\frac{\partial \text{logL}}{\partial \theta _2}&=\frac{2 m}{\theta _2}-\frac{m}{\theta _2+1}-s_2
\end{aligned}
\end{equation}
\noindent from above equations
\begin{equation}
\begin{aligned}
\hat{\theta _1}=&\frac{\left(s_1-n\right)+\sqrt{\left(s_1-n\right){}^2+8 n s_1}}{2 s_1} \\
\hat{\theta _2}=&\frac{\left(s_2-m\right)+\sqrt{\left(s_2-m\right){}^2+8 m s_2}}{2 s_2}
\end{aligned}
\end{equation}
\noindent Hence, using the invariance property of the MLE, the maximum likelihood estimator $\hat{R}_{mle}$ of $ R $ can be obtained by substituting $\hat{\theta}_k$   for $k$=1,2 in equation (27).\\
\begin{equation}
\hat{R}_{mle}=1-\frac{\theta_2^2 \left(\theta_1^3+\left(2 \theta_2+3\right) \theta_1^2+\left(\theta _2+1\right)\left(\theta_2+3\right)\theta_1+\theta_2^2+\theta_2\right)}{\left(\theta_1+1\right) \left(\theta_2+1\right)\left(\theta_1+\theta_2\right)^3} \bigg|_{\theta_1=\hat{\theta}_1,\theta_2=\hat{\theta}_2}
\end{equation}

\subsection{Asymptotic Confidence}
\noindent For an estimator \textbf{$\hat{\theta}_k$} to be  asymptotically  efficient for estimating  $\theta_k$ for large samples,  we should have 
$\sqrt{n_k} \left( \hat{\theta}_k-\theta_k\right) \xrightarrow{\textit{D}} N\left(0,I(\theta_k)^{-1}\right)$
\begin{align*}
I(\theta_k)=&-E\left(\frac{\partial ^2\text{logL}}{\partial \theta _k^2}\right)=-E \left(\frac{n_k}{\left(\theta _k+1\right){}^2}-\frac{2 n_k}{\theta _k^2}\right)=\frac{2 n_k}{\theta _1^2}-\frac{n_k}{\left(\theta _1+1\right){}^2} \\
I_1=I(\theta_1)=&-E \left(\frac{n}{\left(\theta _1+1\right){}^2}-\frac{2 n}{\theta _1^2}\right)=\frac{2 n}{\theta _1^2}-\frac{n}{\left(\theta _1+1\right){}^2} \\
I_2=I(\theta _2)=&-E \left(\frac{m}{\left(\theta _2+1\right){}^2}-\frac{2 m}{\theta _2^2}\right)=\frac{2 m}{\theta _2^2}-\frac{m}{\left(\theta _2+1\right){}^2}\\
\end{align*}
where     $n_1 = n$ and      $ n_2= m$\\
\noindent therefore, as  $n\rightarrow \infty$ and $m  \rightarrow \infty$ 
\begin{equation*}
\frac{\hat{R}-R}{\sqrt{\frac{d_2^2}{m I_2}+\frac{d_1^2}{n I_1}}}\overset{D}{\longrightarrow }N(0,1)
\end{equation*}
where
\begin{align*}
d_1=&\frac{\partial R}{\partial \theta_1}=\frac{\theta_1 \theta_2^2 \left(\theta_1^3+2 \left(\theta _2+3\right) \theta_1^2+\left(\theta_2^2+8 \theta_2+12\right) \theta_1+2 \left(\theta_2^2+3\theta _2+3\right)\right)}{\left(\theta_1+1\right)^2 \left(\theta_2+1\right) \left(\theta_1+\theta_2\right){}^4} \\
d_2=&\frac{\partial R}{\partial \theta_2}=-\frac{\theta_1^2 \theta_2 \left(\theta_2^3+6 \theta_2^2+12 \theta_2+\theta_1^2 \left(\theta_2+2\right)+2 \theta_1 \left(\theta _2^2+4\theta _2+3\right)+6\right)}{\left(\theta_1+1\right) \left(\theta _2+1\right)^2 \left(\theta_1+\theta _2\right)^4}
\end{align*}

Interval estimators and 100$(1-\alpha)\%$ confidence interval for R can be obtained by using the asymptotic distribution of $\hat{R}$, and obtained as
\begin{equation*}
\hat{R}\pm Z_{\frac{\alpha }{2}} \sqrt{\frac{\hat{d_2^2}}{m\hat{I}_2}+\frac{\hat{d_1^2}}{n \hat{I}_1}}
\end{equation*}

\section*{Conclusion:}
We have proposed the new distribution Lindley-Exponential (L-E) distribution generated by Lindley distribution. We have derived important properties of the L-E distribution like moments, entropy, asymptotic distribution of sample maximum and sample Minimum. We have illustrated the application of L-E distribution to two real data sets used by researchers earlier. By comparing L-E distribution with other popular models we conclude that L-E distribution performs satisfactorily or better.

\pagebreak

\section{Appendix}
\subsection{A.1- Dataset used in Illustration 1:}
\begin{small}

\begin{table}[htbp]
  \centering
  \caption{The remission times (in months) of bladder cancer patients}
    \begin{tabular}{|r|r|r|r|r|r|r|r|r|r|r|r|r|} \hline
    0.08  & 2.09  & 3.48  & 4.87  & 6.94  & 8.66  & 13.11 & 23.63 & 0.2   & 2.23  & 0.26  & 0.31  & 0.73 \\ \hline
    0.52  & 4.98  & 6.97  & 9.02  & 13.29 & 0.4   & 2.26  & 3.57  & 5.06  & 7.09  & 11.98 & 4.51  & 2.07 \\ \hline
    0.22  & 13.8  & 25.74 & 0.5   & 2.46  & 3.64  & 5.09  & 7.26  & 9.47  & 14.24 & 19.13 & 6.54  & 3.36 \\ \hline
    0.82  & 0.51  & 2.54  & 3.7   & 5.17  & 7.28  & 9.74  & 14.76 & 26.31 & 0.81  & 1.76  & 8.53  & 6.93 \\ \hline
    0.62  & 3.82  & 5.32  & 7.32  & 10.06 & 14.77 & 32.15 & 2.64  & 3.88  & 5.32  & 3.25  & 12.03 & 8.65 \\ \hline
    0.39  & 10.34 & 14.83 & 34.26 & 0.9   & 2.69  & 4.18  & 5.34  & 7.59  & 10.66 & 4.5   & 20.28 & 12.63 \\ \hline
    0.96  & 36.66 & 1.05  & 2.69  & 4.23  & 5.41  & 7.62  & 10.75 & 16.62 & 43.01 & 6.25  & 2.02  & 22.69 \\ \hline
    0.19  & 2.75  & 4.26  & 5.41  & 7.63  & 17.12 & 46.12 & 1.26  & 2.83  & 4.33  & 8.37  & 3.36  & 5.49 \\ \hline
    0.66  & 11.25 & 17.14 & 79.05 & 1.35  & 2.87  & 5.62  & 7.87  & 11.64 & 17.36 & 12.02 & 6.76  &  \\ \hline
    0.4   & 3.02  & 4.34  & 5.71  & 7.93  & 11.79 & 18.1  & 1.46  & 4.4   & 5.85  & 2.02  & 12.07 &  \\ \hline
    \end{tabular}%
  \label{tab:addlabel}%
\end{table}%
\end{small}

\subsection{A.2- Dataset used in Illustration 2:}
\begin{small}
\begin{table}[htbp]
  \centering
  \caption{Waiting times (min.) of 100 bank customers}
    \begin{tabular}{|r|r|r|r|r|r|r|r|r|r|} \hline
    0.8   & 0.8   & 1.3   & 1.5   & 1.8   & 1.9   & 1.9   & 2.1   & 2.6   & 2.7 \\
    2.9   & 3.1   & 3.2   & 3.3   & 3.5   & 3.6   & 4     & 4.1   & 4.2   & 4.2 \\
    4.3   & 4.3   & 4.4   & 4.4   & 4.6   & 4.7   & 4.7   & 4.8   & 4.9   & 4.9 \\
    5.0   & 5.3   & 5.5   & 5.7   & 5.7   & 6.1   & 6.2   & 6.2   & 6.2   & 6.3 \\
    6.7   & 6.9   & 7.1   & 7.1   & 7.1   & 7.1   & 7.4   & 7.6   & 7.7   & 8 \\
    8.2   & 8.6   & 8.6   & 8.6   & 8.8   & 8.8   & 8.9   & 8.9   & 9.5   & 9.6 \\
    9.7   & 9.8   & 10.7  & 10.9  & 11.0  & 11.0  & 11.1  & 11.2  & 11.2  & 11.5 \\
    11.9  & 12.4  & 12.5  & 12.9  & 13.0    & 13.1  & 13.3  & 13.6  & 13.7  & 13.9 \\
    14.1  & 15.4  & 15.4  & 17.3  & 17.3  & 18.1  & 18.2  & 18.4  & 18.9  & 19.0 \\
    19.9  & 20.6  & 21.3  & 21.4  & 21.9  & 23    & 27    & 31.6  & 33.1  & 38.5 \\ \hline
    \end{tabular}%
  \label{tab:addlabel}%
\end{table}
\end{small}
\end{document}